\documentclass[12,pt]{article}
\usepackage{amsfonts,amssymb,amsmath,latexsym}
\newcommand{\be}{\begin{equation}}
\newcommand{\ee}{\end{equation}}
\newcommand{\bea}{\begin{eqnarray}}
\newcommand{\eea}{\end{eqnarray}}
\newcommand{\ba}{\begin{array}}
\newcommand{\ea}{\end{array}}
\newcommand{\bt}{\begin{tabular}}
\newcommand{\et}{\end{tabular}}

\newcommand{\fr}{\frac}
\newcommand{\ci}{\cite}
\newcommand{\cl}{\centerline}
\newcommand{\bs}{\bigskip}
\newcommand{\vs}{\vspace}

\newcommand{\en}{\eqno}

\newcommand{\bbib}{\begin{thebibliography}}
\newcommand{\ebib}{\end{thebibliography}}

\newcommand{\mbb}{\mathbb}

\newcommand{\und}{\underline}
\newcommand{\np}{\newpage}

\begin{document}
\titlepage
\cl{\bf VAN DER WAALS $\sigma$-MODEL}
\cl{\bf AND TOPOLOGICAL EXITATIONS }
\cl{\bf WITH LOGARITHMICAL  ENERGY.}
\vs{1cm}
\cl{\bf S.A.Bulgadaev}
\vspace{0.5cm}
\centerline{L.D.Landau Institute for Theoretical Physics}
\centerline{Kosyghin Str.2, Moscow, 117334, RUSSIA}
\vs{1cm}
\centerline{A talk given at seminar}
\bs
\cl{\large "Properties and Dynamics of Defects}
\bs
\cl{\large in Liquid Crystals"}
\bs
\cl{23 August 1999, MPIPKS, Dresden, Germany}

\newpage
\cl{{\bf I. INTRODUCTION}}

\bs

The stable topological  excitations (TE)
can exist in systems with degenerate minima, which form some manifold
${\cal M}$ with  nontrivial topology.
Such TE take place in many physical systems, in particular,
in superfluid $He^3$ and $He^4$, in different liquid crystals, in magnets.
The two main problems are connected with the existence of TE:

1. Their classification and properties.

2. Their influence on the properties of the systems.

The most important  property of the TE is connected with their influence
on the behaviour of correlations in the systems.
In general, this influence depends on interrelations between such fundamental
properties of the systems as symmetry and topology.
For example, the influence of the TE is especially strong in systems with
initial scale invariance, since
the TE introduce in the theory a new effective mass scale \ci{1}
$$
m \sim a^{-1}\exp (-{\cal S}_{TE}),
\en(1)
$$
where $a$ is some UV cut-off parameter (a core radius or a lattice constant)
and  ${\cal S}_{TE}$ is the dimensionless "energy" (or action) of the TE.
In the cases,
when one has many TE, they can form some textures,
ordered or disordered, which also influence on behaviour of the systems.
In these cases the influence depends also on the interaction between TE,
which, in its turn, is determined again by interrelations between
symmetries and geometry of the systems. If an interaction of TE is
strong enough,

\und {it can induce the topological phase transition (TPT)
in the system of TE,}

\und {which can drastically change a character of correlations in system.}

\noindent This effect, connected with TPT, is well studied in
low-dimensional ($D \le 2$) systems [2-4,5-9].
In order to understand its main
reasons let us resume properties of such low-dimensional systems.

\bs

\cl{\bf II. THE TE WITH LOGARITHMIC ENERGY}

\cl{\bf IN LOW-DIMENSIONAL NS-MODELS}

\bs

For investigation of the topological properties it is convenient to consider
nonlinear $\sigma$-models (NSM) on the spaces ${\cal M}$, which are the
long-wave  approximation of the corresponding Ginzburg - Landau (GL) type
theories. The topological properties of the space ${\cal M}$ are described
by the homotopic groups $\pi_i({\cal M}), \, i= 0,1,...$

\bs

1. \und {$D=2.$}

\bs

The most popular two-dimensional NSM are the next:

\bs

\underline {1. NS-model on a circle $S^1$}, which is defined by the action
\ci{10}
$$
{\cal S} = \fr{1}{2\alpha} \int d^2 x |\partial \psi |^2, \quad
\psi = \exp (i\phi) \in S^1.
\en(2)
$$
The relevant homotopical groups are
$$\pi_i(S^1) = 0, \, i\ne 1, \quad \pi_1(S^1) = \mbb {Z}.
\en(3)
$$
Due to this, only one type of TE, the vortices, exist in this model.
Under a vortex one can understand any TE (with a nontrivial vorticity or
pure potential), corresponding to $\pi_1(S^1) = \mbb {Z}.$
The energy of one  vortex  with topological charge
$e\in \mbb {Z}$ is logarithmically divergent
$$
E= \fr{e^2}{2\alpha} 2\pi \ln \fr{R}{a},
\en(4)
$$
where $R$ is a space radius. The energy of $N$-"vortex" solution, $E_N,$
with the full topological charge $e = \sum _{i=1}^N e_i = 0$ is finite
and equals
$$
E_N= \fr{2\pi}{2\alpha}\sum_{i\ne k}^N e_i e_k
\ln \fr{|{\bf x}_i-{\bf x}_k|}{a} + C(a) \sum_i^N e_i^2,
\en(5)
$$
here $C(a)$ is some nonuniversal constant, determining "self-energy"
(or a core energy) of vortices and depending on type of a core regularization.
From (5) it follows that vortices interact through logarithmic potential.
Just this interaction of vortices induces the TPT in all 2D systems
with continuous abelian symmetry, described  by this NSM [5-7].

\bs

\und {2. NS-model on a sphere $S^2$} is defined by action \ci{11}
$$
{\cal S} = \fr{1}{2\alpha} \int d^2 x (\partial {\bf n} )^2, \quad
{\bf n}  \in S^2.
\en(6)
$$
The relevant homotopical groups are
$$
\pi_i(S^2) = 0, \, i < 2, \quad \pi_2(S^2) = \mbb {Z}.
\en(7)
$$
There is also only one type of TE, the instantons,
with topological charges $q \in \mbb {Z}$ equal to the degree of
the corresponding mapping $S^2 \to S^2.$
The energy of $N$ instantons with topological charges $q_i, \, i = 1,..N$
$$
E_N = \fr{4\pi}{\alpha} \sum_{i=1}^N |q_i|
\en(8)
$$
Thus, the instantons do not interact and only weak, dipole-dipole like,
interaction exist between instantons and anti-instantons \ci{10}.
The different types of interactions of TE in these models is determined by
different character of TE, corresponding to groups $\pi_1(S^1)$ and
$\pi_2(S^2),$ in $D=2.$

\bs

In 2D the TE, described by $\pi_1(S^1)$, correspond to the

\und {open space boundary and have logarithmic energy,}

\bs

\noindent while the TE, described by $\pi_2(S^2)$, correspond to the boundary
shrinked into one point and for this they have finite energy.

\bs

In principle, in both models the TE of vortex and instanton types
are possible.
In the first model the "neutral" configurations of vortices also
correspond to
the shrinked space boundary and could be classified by $\pi_2(S^1).$
But, since $\pi_2(S^1)=0,$ all neutral configurations belong to one type.
Analogously, in the second model the vortices (or merons) are possible.
Due to $\pi_1(S^2)=0,$ they are unstable. Their "neutral",
dipole-like, configurations can have different topological structure,
corresponding to $\pi_2(S^2)= \mbb {Z},$ and are stable.

Analogous situation takes place in

\bs

2. \und {1D systems  with long-range  interaction of type}
$$
J(r) \sim 1/r^2.
\en(9)
$$
This form of $J(x)$ is determined by the
\und {scale (even conformal) invariance} \ci{12}.

The corresponding \und {1D NS-models} have the next Hamiltonian
\bs
$$
{\cal H} = \fr{1}{2\alpha}
\int dx dx' J(x-x')({\bf s}(x) -{\bf s}(x'))^2.
\en(10)
$$
\bs
Here, due to $D=1$, the important manifolds and homotopical groups are
$$
{\cal M} = \{{\bf s}_i\}, \, i=1,...p, \,
\quad \pi_0(\{{\bf s}_i\}) \ne 0, \quad \pi_i = 0, \, i \ge 1;
$$
$$
{\cal M} = S^1, \quad \pi_1(S^1) =\mbb {Z},
$$
where ${\bf s}$ is a classical spin and  a number of discrete spin states
$p$ can be finite or infinite \ci{2,3,4}.
To the discrete set $\{{\bf s}_i\}$ and the open boundary correspond
domain walls or kinks, which connect different minima, and to the
compactified space $\mbb {R}^1$, i.e. $S^1,$ correspond 1D instantons,
which are really  the vortices, since they
correspond to $\pi_1(S^1)$ \ci{12,13}.

\bs

\noindent \und {In 1D case the kinks described by $\pi_0(\{{\bf s}\})$
have logarithmic energy}

\noindent \und {and can induce TPT}, while instantons have finite energy
and do not interact between themselves.

\bs

\und {\bf Resume}:

\bs

In low-dimensional systems ($D\le 2$) the TE with logarithmic energy exist
in systems described by

\bs

1) \und {\bf conformal invariant NS-models,}

\bs

2) \und {\bf defined on spaces with discrete abelian group $\pi_{D-1}\ne 0.$}

\bs

These TE can induce TPT, which describe such important physical phenomena
as localization-delocalization \ci{14}, coherence-decoherence \ci{15},
melting \ci{8} and others.
All these transitions change a character of correlations in systems.

\bs

Then, the two main
\und {\bf Questions} appear:

\bs

1) Can such TE exist  in higher-dimensional systems?

\bs

2) Can they influence on correlations so strong?

\bs

In this talk  we propose one model which helps us to answer on the
first question
and can help answer on the second question.

\bs

\cl{\bf III. 3D VAN DER WAALS NS-MODEL ON $S^2$}

\bs

1. \und {Motivation.}

\bs

An investigation of the possibility of existence of the TE with logarithmic
energy in 3D scale invariant systems and their influence on correlations.

\bs

2. \und {Model.}

\bs

Let us consider 3D lattice with the order parameter (OP) ${\bf n}$ in each
lattice site. The OP can be:

1) a unit vector ${\bf n}, \quad n^2 =1, \quad {\bf n} \in S^2,$

it can represent a magnet;

2) a unit rod or a director  ${\bf n} \in S^2/Z_2 = RP^2,$

it can represent a liquid crystal or molecular crystal.

\bs

Since the relevant homotopy groups are \ci{16}
$$
\pi_2(S^2) = \pi_2(RP^2) = \mbb {Z},
\en(11)
$$
$$
\pi_3(S^2) = \pi_3(RP^2) = \mbb {Z},
\en(12)
$$
it will be more convenient to consider the OP ${\bf n} \in S^2.$
All results will takes place with non-essential modifications for
${\bf n} \in RP^2$  too.
Due to its vectorness,   the OP can interact by different type
of interactions:

1) exchange type  $\sim ({\bf n}_r \cdot {\bf n}_{r'}) V(r-r')$,

2) dipole-like  $\sim  (n_r^i D_{ik}(r-r') n_{r'}^k ) V(r-r')\quad V(r) \sim 1/r^3$,

$D_{ik}(r) = \delta_{ik} - 3\fr{x^i x^k}{x^2}$,

3) van der Waals type $V_{vdW}(r) \sim 1/r^6$.

All they can be represented in the next form
$$
{\cal E} \sim (n^i_r D_{ik} n^k_{r'}) V(r-r'),\quad
V(r) \sim 1/r^{\sigma},
\en(13)
$$
where  $\sigma$ defines an asymptotic behaviour of the potential $V(r).$

As a first approximate attempt to the problem one can compose from all
these types of interaction the simplified one, which must conserve
the two main properties:

1) a scale invariance of the corresponding Hamiltonian ${\cal H},$

2) a vectorness of the OP.

In result one gets \und {the lattice vector van der Waals model}
with ${\cal H}$
$$
{\cal H} = - \fr{J}{2} \sum_{r \ne r'} ({\bf n}_r \cdot {\bf n}_{r'})V_{vdW}(r-r').
\en(14)
$$
The analogous approximation, for example, was used by Nelson \ci{8}
in the theory of 2D melting.

In long-wave continuous approximation our model passes into the

\und {vector van der Waals NS-model} with a partition function
$$
{\cal Z}_{vdW} =\int D{\bf n} e^{- {\cal S}_{vdW}[{\bf n}]},
$$
$$
{\cal S}_{vdW}[{\bf n}] =  - \fr{1}{2\alpha}\int d^3x d^3x'
({\bf n}(x){\bf n}(x'))V_{vdW}(x-x'),
\en(15)
$$
$$
V_{vdW}(x)= \left.\int \fr{d^3k}{(2\pi)^3} e^{i({\bf k}{\bf x})}|k|^3 f(ka)
\right|_{|x| \gg a}
\sim 1/|x|^6
\en(16)
$$
where
$${\bf n}^2 =1, \quad \alpha \sim  1/J\beta,
\en(17)
$$
$f(ka)$ is a regularizing function with next asymptotics
$$
f(ka)_{ka \ll 1} \simeq 1+O(ka), \quad
f(ka)_{ka \gg 1} \to 0.
\en(18)
$$
From now on we omit an index vdW vor brevity.
A scale invariance of the model at large distancies follows
immediately from large-distance asymptotics of $V(x)$ and
dimensionlessness of the OP ${\bf n}.$ Moreover, ${\cal S}$ is
conformal invariant at large distancies, i.e. it is invariant
under conformal transformations:
$$
x_i \to  x'_i = x_i/x^2, \quad x \to x' = 1/x, \quad x_i/x = x'_i/x',
$$
$$
d^3 x \to d^3 x/|{\bf x}|^6,\quad
\fr{1}{|{\bf x}_1-{\bf x}_2|^{6}} \to
\fr{|{\bf x}_1|^6 \;|{\bf x}_2|^6}{|{\bf x}_1-{\bf x}_2|^{6}},
\en(19)
$$
and, consequently,
$$
{\cal S} \sim \int d^3 x_1 d^3 x_2 \;
\fr{({\bf n}_1{\bf n}_2)}{|{\bf x}_1-{\bf x}_2|^{6}} \to {\cal S}
$$
For this reason  this model can be  named also the
\und {3D conformal NS-model} \ci{17}.
\np
The corresponding Euler - Lagrange equation has a form
$$
\int V(x-x') {\bf n}(x')d^3 x' -
{\bf n}(x) \int ({\bf n}(x){\bf n}(x')) V(x-x') d^3 x' = 0.
\en(20)
$$
The Green function $G(x)$ of the conformal kernel
$V(x)$ we define by next equation
$$
\int V(x-x'') G(x''-x') d^3 x'' = \delta (x-x')
\en(21)
$$
It has the following form
$$
G(x) = \left.\int \fr{d^3 k}{(2\pi)^3}\fr{e^{i({\bf k}{\bf x})}}
{k^3 f(ka)} \right|_{r \gg a} \simeq
- \fr{2\pi}{(2\pi)^3 (2)^{1/2}
\Gamma(3/2)} \ln (r/R)
\en(22)
$$
and logarithmic asymptotic behaviour.

The action (15) can be represented in a form, admitting a small deviation
expansion of ${\bf n}$
$$
{\cal S}[{\bf n}] =  \fr{1}{4\alpha}\int d^3x d^3x'
({\bf n}(x) - {\bf n}(x'))^2 V(x-x').
\en(23)
$$
In this case in its Fourier form can appear a usual, $\sim k^2,$ term
which breaks a scale invariance. Then one can consider a more general
model, including a local gradient term in an explicit form
$$
{\cal S}_{loc} = \fr{1}{2\alpha'} \int d^3x (\partial {\bf n})^2.
\en(24)
$$
This term is not scale invariant in $3D$ and for this reason the whole
action
$$
{\cal S}_{tot} = {\cal S} + {\cal S}_{loc}
\en(25)
$$
is also not scale invariant. Its form in Fourier space will be
$$
{\cal S}_{tot} = {\cal S}_{loc} + {\cal S} = \int \fr {d^3 k}{(2\pi)^3}
|n(k)|^2 \left((\fr{1}{2\alpha'}+ C)k^2 +  \fr{1}{2\alpha} k^3 +
...\right),
\en(26)
$$
A similar action was obtained earlier in thve theory of liquid
crystals \ci{18}.
If in the system there is a point,
where a "rigidity" $(\fr{1}{2\alpha'}+ C)= 0$,
then in this point one obtains  a scale invariant action (15) with a
first term $\sim k^3$. From a point of view of the GL theory this is
similar to the
\und {tricritical point}, where a term $\sim \psi^4$ is absent in the
expansion of effective potential of the theory on nonlinearities \ci{19}.
\np

\cl{\bf IV. THE TE WITH LOGARITHMIC ENERGY}

\cl{\bf IN 3D VAN DER WAALS NS-MODEL ON $S^2$}

\bs

A nontriviality of $\pi_2(S^2) = \mbb {Z}$ means that  in the model
there are point-like  TE, corresponding to the open boundary.
The simplest such  excitation is the "hedgehog". It is a solution of
equation (20) and has the next
asymptotic form
$$
\left.{\bf n}(x)\right|_{|x| \gg a} \simeq  \fr{x^i}{|x|}.
\en(27)
$$
The action ${\cal S}$ of this solution is
$$
{\cal S} = 4\fr{(4\pi)^2}{2\alpha} \fr{4\pi}{(2\pi)^3}
\int \fr{dk}{k} f(ka)
$$
$$
\approx  4\fr{(4\pi)^2}{2\alpha} \fr{4\pi}{(2\pi)^3} \ln (R/a).
\en(28)
$$
Its topological charge $q \in \mbb {Z}$ is the degree of the
corresponding mapping.
The energy of two "hedgehogs" with a full topological charge
$q =0$ is finite and the interaction of two "hedgehogs" with charges
$q_1$ and $q_2$ on large distancies has a form of the Green function
$G(x)$ of the kernel $V(x)$
$$
H_{12} (r) = e_1 e_2 G(r) \simeq - e_1 e_2 \fr{2\pi}{(2\pi)^3 (2)^{1/2}
\Gamma(3/2)} \ln (r/R).
\en(29)
$$
Note that in the usual local 3D NS-model (24) such TE have the energy linear
in $R$
$$
E \sim \fr{R}{\alpha'}.
\en(30)
$$
A nontriviality of another homotopical
group $\pi_3(S^2) = \mbb {Z}$ means that the "neutral"
configurations, having a full topological charge equal to 0 and
corresponding to the shrinked boundary,
can also have different topological structures.
They are characterized by topological invariant, coinciding with
the


\cl{\und {Hopf invariant $H \in \mbb {Z}$}}


\noindent of the corresponding mapping $S^3 \to S^2.$
This invariant is connected  with linking number and can be expressed
through the integrals over $\mbb {R}^3$ or over compactified space
$\mbb {R}^3 \simeq S^3.$ Some its properties are described in Appendix.
Just this invariant is an analog of the topological charge of 2D instantons.
Note, that this additional topological invariant classifies "neutral"
configurations in all 3D NS-models defined on sphere $S^2,$ in particular,
in the usual local model (24).
Thus, the "hedgehog" excitations in 3D van der Waals NSM  have
properties reminiscent of the mixed properties of the two-dimensional
vortices and instantons:

1) \und { their topology is described by $\pi_2(S^2)$,}

\und {but they interact as vortices through logarithmic potential;}

2) \und { their "neutral" configurations are classified by integer topological}

\und {Hopf invariant $H$.}

\bs

\cl{\bf V. DISCUSSION, APPLICATIONS}

\bs

1. \und { A possibility of TPT}

\bs

The TE with logarithmic energy can induce TPT in system of such TE.
Simple arguments by Kosterlitz and Thouless show this for any dimensional
case. Let us admit that in D-dimensional system there is such TE with energy
$$
E = A \ln R/a.
$$
Consider the corresponding free energy $F$ per one such TE
$$
F = E - TS, \quad
$$
where T is a temperature, $S$ is an entropy.  One can estimate  the entropy
of the TE as a logarithm of the number of possible places of the TE in space
$$
S= k_B\ln (R/a)^D = k_B D \ln (R/a).
$$
Then one has for free energy of one TE
$$
F = E - TS = (A - k_B D T) \ln (R/a)
$$
Free energy $F$ becomes equal to 0 and  changes it sign at
$$
T_{KT} = A/k_B D
$$
It means that at  $T > T_{KT}$ it becomes energetically favorable
to birth such TE and they can generate spontaneously, while at
$T < T_{KT}$ one needs positive free energy to birth these TE. More
detailed study of this transition in low-dimensional systems has shown
that these arguments correspond to the first order approximation in
the renorm-group (RG) approach.  In higher orders of RG  the contributions,
taking into account more detailed information about geometry of space
${\cal M},$ type of the corresponding topological charges and
their interaction, appear in RG equations. In principle, than more
complicated topological structures appear, then different complications,
destroying this TPT,  become more probable.

In 3D van der Waals NS-model there are some  additional complications,
which can remove the TPT or complexify its study:

\bs

1) additional topological invariant, the Hopf invariant $H$,
which distinguishes different "neutral" configurations;

\bs

2) one needs to conserve a condition, equalizing a rigidity to 0,
in order  a logarithmic interaction can not  be screened
during renormalization;

\bs

3) a conformal symmetry, which is, on our opinion, a hidden reason of
existence of TPT in low-dimensional systems,
is finite in spaces with $D>2,$  while in spaces with $D \le 2$ it
is infinite-dimensional;

\bs

4) other, not so evident, complications.

\bs

2. \und {Generalizations.}

\bs

\und {1. D-dimensional generalization.}

\bs

The first condition from the resume defines form of action $S$ in any dimension
and the second one defines partially a topology of ${\cal M}$
$$
S=\frac{1}{2\alpha}
\int d^D x d^D x' \psi_a (x)\boxtimes_{ab}^{(D)}(x-x')\psi_b (x'),
$$
where $\psi \in {\cal M}, \; a,b = 1,2...,n, \; n$ is a dimension of
${\cal M}$ and
a form of kernel $\boxtimes$ depends on dimension of space $D$.
For decoupled  internal and physical spaces
$\boxtimes$ can be decomposed
$$
\boxtimes_{ab}^{(D)}(x)= g_{ab} \Box_D (x),
$$
where $g_{ab}$ is the Euclidean  metric
of the space $\mathbb{R}^{N(n)}$, in which a manifold
${\cal M}$ can be embedded.  In the momentum space, for small $k$
$$
\Box_D (k)\simeq |k|^D (1 + a_1 (ka) + ...),
$$
where $a$ is a UV cut-off parameter.
Action ${\cal S}$ can be named $D$-dimensional {\emph{conformal}}
nonlinear $\sigma$-model. The kernel $\Box_D$ generalizes
an usual  local and conformal kernel of two-dimensional
$\sigma$-model
$$
\Box(k)\equiv \Box_2(k) = k^2 .
$$
For local models an expression for $\boxtimes$ can be defined in
terms of manifold ${\cal M}$ only
$$
\boxtimes =  g_{ab}(\phi)\Box \delta(x)
$$
In odd dimensions $\Box_D$ is nonlocal
$$
\Box_D \sim 1/x^{2D}.
$$
The simplest spaces with properties, satisfying the second condition
of the resume, are the spheres $S^{D-1}$.
An existence of topological excitations with logarithmic energy
follows from the invariance of the kernel $\Box_D$ and
the simplest topologically nontrivial
excitations ${\bf n}_i = x_i/r$ under scale and conformal transformations
$$
x_i \to  x'_i = x_i/r^2, \quad r \to r' = 1/r, \quad x_i/r = x'_i/r',
$$
$$
d^D x \to d^D x/|{\bf x}|^{2D},\quad
\fr{1}{|{\bf x}_1-{\bf x}_2|^{2D}} \to
\fr{|{\bf x}_1|^{2D} \,|{\bf x}_2|^{2D}}{|{\bf x}_1-{\bf x}_2|^{2D}},
$$
and, consequently,
$$
{\cal S} \sim \int d^D x_1 d^D x_2 \; \fr{({\bf x}_1{\bf x}_2)}{r_1\;r_2}
\fr{1}{|{\bf x}_1-{\bf x}
_2|^{2D}}
$$
is invariant and dimensionless.
By direct calculation one can show that
$$
{\cal S} [{\bf n}] \sim  C_D \ln \fr{R}{a},
$$
and the different excitations interact
through potential $G_D(r),$  inverse to $\Box_D(r),$ which has a
logariphmic behaviour at large distancies
$$
G_D(r) \sim \ln \fr{r}{R}.
$$
Just these conformal kernels and their logarithmic Green functions have
appeared
under consideration  of logarithmic gases and equivalent field theories
in \ci{20}.

\bs

2. \underline {Multicomponent generalization for $D>2$}.

\bs

A simple generalization of the sphere $S^{D-1}$, analogous to the torus
$T^n$ in 2D case, is a bouquet of $n$ spheres
$$
B_n^{D-1} = S^{D-1}_1\vee... \vee S^{D-1}_n
$$
and all spaces ${\cal M}$ with this first nontrivial topological cell
complex. But, the  vector topological charges, corresponding to different
spheres
will not interact as in a case of torus \ci{21}.
For obtaining an interaction of topological charges in 3D case one
needs to consider NS-models on deformed $B^{D-1}_n$. In particular,
3D van der Waals NS-models on the maximal flag spaces $F_G=G/T_G$
of the simple compact groups $G$, with $\pi_2(F_G)= \mathbb {L}_v,$
(note that a sphere $S^2$ is a particular case of $F_G$:
$S^2 = SU(2)/U(1)$)
will also have topological excitations with interacting vector
topological charges ${\bf Q} \in \mathbb{L}_v$ and  logariphmic energy.
Since $\pi_3(F_G)= \pi_3(G)= \mbb {Z},$
in this case the "neutral" configurations will also have different
topological structures described by group $\pi_3(F_G).$
Thus, in these models the TE will have again the mixed properties of the
two-dimensional vortices and instantons:
their vector topological charges, connected with $\pi_2(F_G)$,
will interact logarithmically as vortices and their "neutral" configurations
will have additional topological structure.

Another interesting possible generalization is the 3D conformal NS-models
on compact  groups $G.$ Since $\pi_2(G) = 0, \, \pi_3(G) = \mbb {Z}$,
then
the instanton-like TE can only  exist in these models.

\bs

\und {3. Applications.}

\bs

The van der Waals NSM, by its construction, is a simplified model
of real systems, consisting from the rod-like molecules, interacting
through the van der Waals potential. But, since

\noindent \und {topological characteristics depend on rough
qualitative properties,}

\noindent \und {not on some inessential details,}

one can hope that

\noindent \und {the proposed van der Waals model can describe the qualitative}

\noindent \und {properties of some real systems.}

An application of ideas, developed under
investigation of this model, to the more
realistic, taking into account an anisotropy of the liquid crystals,
model from \ci{18} may be  especially interesting.

\bs

\cl{{\bf CONCLUSIONS}}

\bs

1. The TE with logarithmic energies have the most strong influence on
correlations in low-dimensional systems with conformal invariance.
They correspond to the open boundary and to group $\pi_{D-1}({\cal M})$.

\bs

2. 3D Van der Waals $\sigma$-model is conformal invariant.

\bs

3. 3D Van der Waals $\sigma$-model on $S^2$ has the pointlike TE,
the hedgehogs, with logarithmic energy and topological charges $q,$
corresponding to homotopical group $\pi_2(S^2) = \mbb {Z}.$

\bs

4. These TE have the mixed properties of the 2D vortices and instantons.
Their "neutral" configurations have different topological structures
characterized by the Hopf invariant.

\bs

5. The possibility of the TPT in 3D systems induced by these TE is discussed.

\bs

6. A generalization of the van der Waals NS-model on other dimensions
and multicomponent systems is proposed.

\bs

\cl{\bf PROBLEMS,  PERSPECTIVES}

\bs

1. General solutions, their classifications.

\bs

2. Renormalization and TPT.

\bs

3. Generalizations and modifications.

\bs

4. Real systems and experiment.

\bs

\cl{\bf APPENDIX. THE HOPF INVARIANT $H$ AND RELATED FACTS}

\bs

The Hopf invariant $H$ classifies the mappings \ci{16}
$$
S^3 \to S^2 = S^3/S^1.
$$
The last equality  means that these mappings are a projection,
which projects some circles $S^1 \in S^3$ into points $\in S^2$.

\bs

An example of such projection.

\bs

Let us take $S^3$ as a spere in $R^4 \simeq C^2$
$$
\fr{z_1}{(|z_1|^2 + |z_2|^2)^{1/2}}, \quad
\fr{z_2}{(|z_1|^2 + |z_2|^2)^{1/2}}, \quad z_{1,2} \in C^2,
$$
and a sphere $S^2$ as a complex projective space $CP^1,$
\noindent where any pairs of the next type
$$z_2 = C z_1$$
gives the same point (here $C$ is arbitrary complex number).

\bs

The Hopf mapping $f$:  $S^3 \to S^2$  can be defined as
$$
f:(z_1,z_2) \to z_1/z_2
$$
for any $z_2 \ne 0.$ It gives $CP^1$, under this all
$$
\lambda z_1, \lambda z_2,
$$
with $|\lambda| = 1$  gives the same point.

\bs

The unit vector ${\bf n} \in S^2$ under this projection can be defined as
$$
n_1 + i n_2 =\fr{2z_1 z_2^*}{|z_1|^2 + |z_2|^2}, \quad
n_3 = \fr{|z_1|^2 - |z_2|^2}{|z_1|^2 + |z_2|^2}
$$
The Hopf invariant   $H$ is a linking number $\{\gamma_1, \gamma_2 \}$
of two projected circles $\gamma_1, \gamma_2 $ in $S^3$,
corresponding to different points of $S^2$ in general positions,
into which they are projected,
$$
H = \{\gamma_1, \gamma_2 \}
\en()
$$
It can be represented in different integral forms.
In $R^3$
$$
\{\gamma_1, \gamma_2 \} =
\fr{1}{4\pi} \oint_{\gamma_1} \oint_{\gamma_2}
\fr{<{\bf r}_{12} \cdot [d{\bf r}_1 d {\bf r}_2]>}
{|r_1-r_2|^3}.
$$
For simple case of one winding of one circle around another $H = 1.$
If a mapping projects each circle $q_i \;(i=1,2)$ times then $H= q_1 q_2.$

\bs

In $S^3$ it can be also represented as
$$
H = \int_{S^3} \theta \wedge d \theta,
\en()
$$
where 1-form $\theta$ is defined as
$$
d \theta = f^{-1}( d\Omega).
\en()
$$
Here $d \Omega$ is a 2-form or an element of the area of $S^2,$
$f^{-1}$ is a mapping  inverse to the  projection mapping.

\bs

I would like to thank the organizers of the seminar for
the opportunity to give this talk. The conversations with
M.Kleman and M.Monastyrsky were very useful for preparation of this talk.
\end{document}